Saarland University

Center for Bioinformatics

Bachelor's Program in Bioinformatics

# Bachelor's Thesis

# Structural Stability and Immunogenicity of Peptides

submitted by

**Eva Kranz**

on March 15, 2010

Supervisor

Prof. Dr. Hans-Peter Lenhof

Advisor

Anna Katharina Dehof, M.Sc.

Reviewers

Prof. Dr. Hans-Peter Lenhof
Dr. Andreas Hildebrandt

# Statement

I hereby confirm that this thesis is my own work
and that I have documented all sources used.

Saarbruecken, March 15, 2010

Eva Kranz

# Acknowledgements

First of all I would like to express my gratitude to Prof. Dr. Hans-Peter Lenhof, who provided me with a relevant and challenging subject to work on. Whenever needed he was available for guidance and support, including questions unrelated to the topics of this thesis.

Anna Katharina Dehof, M.Sc., was an advisor I would recommend to any other student. She always kept track of things and was of great help in terms of C++ programming and many other questions of detail, while remaining seemingly endlessly patient. Also, she gave me invaluable advice when revising my thesis.

Many thanks go to Dr. Andreas Hildebrandt, who helped me install BALL and kindly shared his profound software and programming knowledge, as well as his deep understanding of the biological processes underlying the subject of this thesis.

I appreciate having been pointed to immunogenically interesting proteins by courtesy of Prof. Dr. Eckart Meese.

Furthermore I thank Karin Jostock from the Center for Bioinformatics' secretariat and examination office. She is the most friendly and helpful secretary I have ever met.

I am indebted to Philipp Kiszka, software developer, and the graphic artists Marcus Fröhner and Robert Jung. Philipp's advice helped me bring the webserver alive and the two designers gave the user interface a pleasant look.

I want to gratefully mention the developers standing behind the open source software I used during the process of working out my bachelor's thesis (Apache, BALL, Collabtive, Open Flash Chart, Open Office Suite – to only mention a few).

Finally this thesis would have been much harder to write without occasional distraction and recreation in form of telephone calls from my grandmother, and it would have been impossible without the support and patience of my fiancé Philipp and my parents.

# Table of Contents

# 1  Introduction

The mammalian immune system is a very complex defense system against diseases. Not only can it protect an organism from pathogens, but also against emerging benign [CZW+05] and even malignant tumor cells [AFD+97]. The ability to identify a biochemical structure's origin is crucial for the function of this powerful, intricate system: it should fight infectious agents coming from outside the body, while the organism's own healthy cells must not be attacked.

In some cases the discrimination of the immune system between *self* and *foreign* does not work properly, which leads to an immune response against constituent parts of the organism. This can be a cause for chronical diseases, some of which are even widespread, e.g. diabetes mellitus type I [Bac94] and Hashimoto's thyroiditis [PFB03], but it may also happen in the development of cancer [ABE+01].
At least in the case of cancer the immune system plays a paradoxical role: it fights the disease, or causes additional complications, like e.g. autoimmunity [DEC06, ETK09].

From this it follows that the understanding of the immune system and in particular the knowledge of the etiology of autoimmunity is a key to the development of better therapies and more effective drugs for certain illnesses. These prospects may be one of the reasons why in recent years the adaptive immune system attracted the attention of bioinformaticians at a progressive rate [KLY06].

Many scientists are interested in this field of research, and so are we.

We want to investigate the role that peptide folding stability plays in peptide immunogenicity. Our approach is based on previous work suggesting a direct relationship between a peptide's stability and its ability to induce an immune response.
Our goal is the development of a plausible energy function modeling protein fragment stability using a force field. The resulting model needs to undergo evaluation, therefore we use a reference dataset. Additionally, we want to make an implementation of this model available to the public via a webserver.

## 1.1 Previous Work

The humoral immune system includes the processes of antibody production and antigen recognition (antigen = *anti*body *gen*erator). These processes are integral parts of the humoral immune response both mainly performed by B lymphocytes, also known as B cells.

Research of Camacho *et al.* focused on the question why and how peptides are able to induce an immune response [CKA08]. This property of *immunogenicity* depends on the presence of recognizable epitopes on the surface of a molecule.

The conventional paradigm of humoral immunity claims that a molecule needs a determined three-dimensional structure to be recognized by B cells. Peptides, being protein fragments consisting of a few residues only, hardly bear any stable folding. However, literature provides evidence that some peptides not only are recognizable antigens to B cells, but also lead to the production of antibodies against corresponding regions of the original protein [MBP03, RVT06].

Camacho *et al.* claim the dependency of peptide immunogenicity on the degree of spatial stability. They state that peptides can be divided into three categories depending on their free folding energy $\Delta G$: (1) immune system contacts with instable fragments ($\Delta G > 8$ kcal/mol) are not followed by any response. (2) Weakly stable peptides ($\Delta G > 0$ kcal/mol) lead to an immune response against their own kind, while (3) relatively stable protein fragments with $\Delta G < 0$ kcal/mol induce antibody production against the fragment itself, and also against peptide-like motifs in corresponding proteins.
The latter kind is the class of peptide we are especially interested in.

While the above results are interesting and raise hope of being the answer to the question what the cause of peptide immunogenicity is, the methods and dataset underlying the findings are afflicted with some shortcomings.

Stability can be measured in different units and its computation can be based on several factors. Camacho *et al.* chose to solely base their conclusions on the root mean square deviation (RMSD) of superimposed snapshots of molecular dynamics simulations (MD simulations). While MD simulations provide some insight into the stability of higher order structures of a protein fragment, the resulting RMSD values (unit: Ångström) are not directly related to the $\Delta G$ values (unit: kcal/mol) used for the above described classification. Moreover, this method is rather time-consuming and thus not recommendable for a widespread use.
These problems could be solved by instead using a criterion with a more direct relation to stability that is additionally faster to compute.

The dataset used in Camacho *et al.*'s examination consists of 10 fragments made up of 18 residues, all derived from one single protein: murine histidyl-tRNA synthetase (HRS). This protein as a model for peptide immunogenicity is a good choice. HRS is known to be involved in the pathogenesis of idiopathic inflammatory myopathy and the anti-synthetase syndrome where it plays a role as an autoantigen [YK02].
The dataset should be broadened in order to obtain results with a high degree of reliability.

## 1.2 Task

It was the aim of this thesis to implement a stability criterion based on energy computations using a force field, and to test the implementation with a large dataset.

Our stability criterion is based on computations with the force field AMBER (Assisted Model Building and Energy Refinement), which has been developed especially for molecular dynamics of biomolecules. We use it to compute the energy of a protein fragment with the original coordinates from the respective Potein Data Bank (PDB) file. Instead of simulating the dynamics of a fragment, we compute the energy resulting from its interactions with itself („*self-energy*") and the energy resulting from its interactions with the rest of the protein. The difference between these two terms („energy difference") allows a conclusion to be drawn about the stability of a peptide with the same amino acid sequence as the fragment. In this way we approximate the peptide's free folding energy.

The dataset is being broadened by including a reference dataset of non-homologous protein domains [TQS+05]. We look at every potentially immunogenic fragment by sliding a window over the protein sequence, instead of picking only some fragments. Additionally our computation considers multiple window lengths, reaching from 8 to 22 residues. These lengths are in accordance with the minimum and maximum length of possible epitopes, since they must fit into the binding pocket of major histocompatibility complex (MHC) molecules. These steps result in an enlargement of the dataset by a factor of about 10,000.

Furthermore we want to investigate how the energies correlate with a peptide's solvent accessible surface (SAS) area [LR71, SR73], and with its secondary structure elements. Thus, we implemented computations of these characteristics. This supplement allows for a better understanding of the dependency of a fragment's energy on its other specific features.

Beyond these improvements and extensions we provide a software for a webserver where interested parties can analyze user-defined proteins with respect to their immunogenic properties.

# 2  Biological Background

## 2.1  Immune System

There are two ways of classifying the constituents of the immune system:
1. Innate immunity or adaptive immunity
2. Surface barriers, cellular components, or humoral components

|  | **Innate Immunity** | **Adaptive Immunity** |
|---|---|---|
| **Surface Barriers** | Mechanical barriers, e.g. skin | - |
| **Cellular Components** | Phagocytosis | Antigen recognition |
| **Humoral Components** | Complement system | Antibody production |

*Table 1: Classification of instances of the mammalian immune system*

The question we deal with in this thesis concerns the humoral immune response and therefore belongs into the context of adaptive immunity. It includes cellular (B cells), as well as humoral components (antibodies).

### 2.1.1  Antibodies

Antibodies, also known as immunoglobulins, are proteins created for antigen recognition. They consist of two identical heavy, and two identical light chains, connected by disulfide bonds, leading to a Y-like structure. The two tips of this formation are *hypervariable regions*: the differences in these antigen binding sites of different antibodies lead to high specificity.

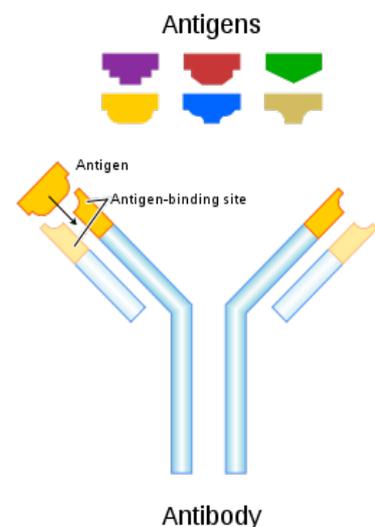

*Figure 1: The light chains (blue, transparent) are shorter than the heavy ones (blue, opaque).*

*Source: Wikimedia Commons*

## 2.1.2  B Cells

B cells evolve from lymphocytes. While antibody assembly is one of their main tasks and the function of interest in our context, they also act as antigen-presenting cells (APCs) and may differentiate further into memory B cells when triggered by the respective signals.

The surface of B cells is covered with B cell receptors (BCRs). They have the same overall structure as antibodies, hence these cells' ability to bind specific antigens.

B cell activation, a necessary step in the differentiation of lymphocytes to B cells, requires certain signals. There are two ways how this process can take place:

During T cell-dependent B cell activation a B cell binds a free antigen (*anti*body *gen*erator) or an antigen presented by an APC, like e.g. a macrophage. When the pathogen cross-links BCRs, the B cell ingests and digests it. The resulting antigen fragments form a complex with major histocompatibility complex (MHC) proteins from inside the B cell on the surface of the cell membrane. Specific T helper cells recognize the antigen-MHC-complex.

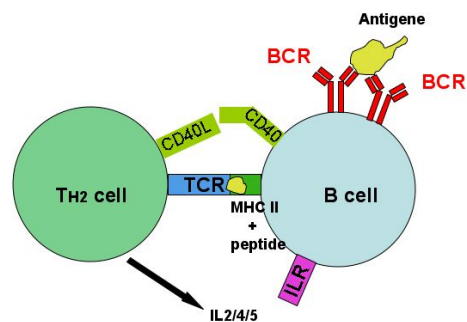

*Figure 2: A T cell (left), B cell (right), and several molecules interact during T cell-dependent B cell activation.*
*Source: Wikimedia Commons*

Finally it comes to a direct interaction between the two cells. The T cell releases effector molecules (cytokines), whereupon the B cell starts proliferation and terminal differentiation into a plasma cell. [JTW+01, Par93]

Some antigens, especially carbohydrates, activate their cognate B cells without additional help from T cells required, which implicates T cell-independent B cell activation. The antigen, which may also be presented by an APC, binds to a particular kind of BCRs, namely IgM antigen receptors, and causes cross-linking. This is sufficient for activating the B cell. [HR09]

### 2.1.3 Epitopes

An epitope, or *antigenic determinant*, is the discriminatory surface structure of a macromolecule that leads to the recognition by the immune system.
In the majority of cases the epitopes of a protein consist of discontinuous amino acids coming together in three-dimensional conformation. These so-called *conformational* epitopes are inevitably broken down upon protein denaturation.

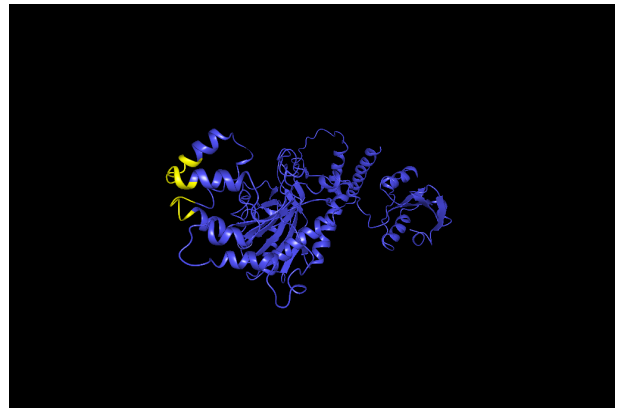

*Figure 3: Sequence-wise separated parts of HRS form a fictional conformational epitope (yellow).*

However, this thesis does not deal with conformational epitopes, but with linear ones only. They consist of about 8 to 22 consecutive residues and thus even occur in peptides.

Epitope length is limited by the spatial prerequisites of MHC binding pockets. MHC class I molecules allow for a length of 8 to 10 residues, while MHC class II molecules present peptides of up to about 22 residues in length. These constraints must be considered when examining potential antigens from protein fragments and it is reflected in our choice of examined window lengths.

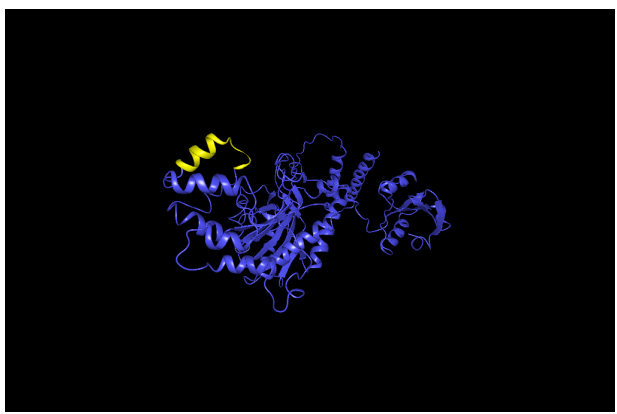

*Figure 4: Consecutive residues of HRS form a fictional linear epitope (yellow).*

## 2.2 Humoral Immune Response

The humoral immune response (lat. *humor* = liquid) takes place in the body fluids blood and lymph. It is the process of antibody production by B cells.

We describe the course of an immune response.

First of all the involved activated B cell recognizes its specific antigen, usually a *non-self* molecule, that is an intruder like e.g. a bacterium, by binding it with its membrane-bound antibodies, the BCRs. This step is called *antigen-recognition*.

The recognition induces a complicated maturation process: the B cell's terminal differentiation into a plasma cell.

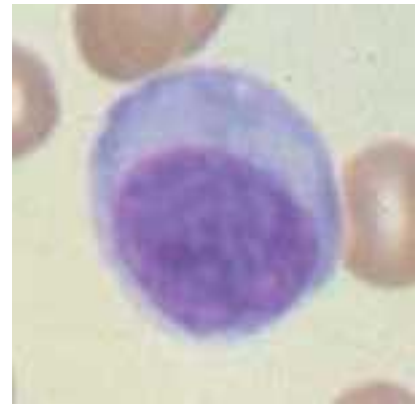

*Figure 5: Plasma cells are the result of terminal differentiation of B cells. The process usually takes place in lymph nodes.*
*Source: Wikimedia Commons*

The resulting cell starts the production of large volumes of antibodies against the antigen that triggered the immune response. The immunoglobulins are being secreted into the body fluids.

Now it comes to an *antigen-antibody reaction*.
Since each antibody can bind two antigens and some antigens have more than one epitope on their surface, a so-called *immune complex* is being formed, consisting of several antibodies and antigens. Binding the antigen implicates its immunization.
In the final step of the immune response the immune complex prepares the antigen for degradation. This process can be realized through phagocytosis or by the complement cascade, for instance.

## 2.3 Immunogenicity

The immunogenicity of a protein depends on its unique folding because the antigen-presentation, which is a precondition for the antigen-recognition by B cells, is controlled by the folding stability [TMD+04]. When the protein gets denatured, most of its epitopes will get lost because they lose their stable conformation. Thus, antigen-stability is possibly one of the crucial points of immunogenicity.

This leads to the question why few peptides, though lacking a noteworthy stability, not only lead to antibody production against their own structure, but even induce an immune response against motifs similar to the peptide, which occur in proteins.

The answer suggested by Camacho *et al.* proposes a dependency of a peptide's immunogenicity on its stability, however marginal.

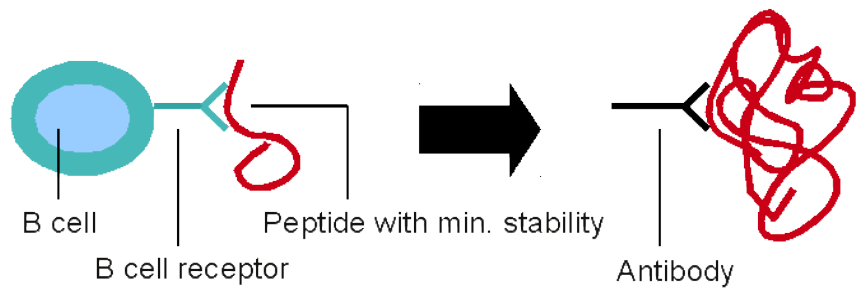

*Figure 6: According to Camacho* et al. *a peptide with minimal stability induces antibody production against the corresponding part of the cognate protein.*

When a peptide leads to the production of antibodies even recognizing the corresponding structure in the native protein, this part of the protein is likely to be comparatively stable.

In order to estimate the immunogenicity of a peptide we compute the stability of the corresponding residues in the full protein.

In this way we make sure that the calculation is based on a reasonable start conformation: peptides are often protein fragments resulting from cleavage by a protease. Hence, it can be assumed that the start conformation of the peptide is about the same as the conformation of the corresponding protein residues.

# 3 Implementation

In the following we describe with which software and how exactly we implemented the program structure.

## 3.1 Choice of Software

For the implementation of fragment energy, SAS area, and secondary structure computation we chose to use a C++ application framework. The Biochemical Algorithms Library BALL [BKL99, Koh01] provides classes and methods for the import and export of PDB files, it supports the analysis and comparison of protein features, and it implements several force fields, e.g. AMBER, including methods for solvation [SRD08].

## 3.2 Program Flow

### 3.2.1 Protein Preparation

First, we explain the expected command line arguments and the file import.
The C++ program is given the path to a PDB file, a name for the output folder, and a range of window lengths. The range is specified by the minimum and maximum window length to compute. With regard to the capacity of MHC molecules' binding sites we recommend a standard minimum length of 8, and a standard maximum length of 22 residues.
The program imports the PDB file and reads it into a *system*, the corresponding data structure.

In a second step, the program applies a special treatment on chains and ligands.
Many proteins consist of more than one chain. In this step the protein is split up into its chains, each of which is treated as a system itself in the following iteration.

We are not interested in ligands which are part of a PDB file, so we delete them. If a chain does not contain anything but a ligand, it will be empty after the deletion. We do not consider it any further, for that reason we just jump to the next chain.

The final preparation step deals with atom names and hydrogen atoms.
Atom names from PDB files sometimes need normalization. The correct names are being retrieved from entries of a fragment database.
These entries also allow for adding missing hydrogen atoms. PDB files come without them since the methods used for protein structure determination are not sensitive enough to detect these smallest atoms.
Subsequently the missing bonds between the new atoms and the protein are being added.
After having checked the consistency of our model, we optimize the hydrogen atom positions: an AMBER force field is being set up and we perform 50 steps of energy optimization using a conjugate gradient minimizer. The changes of the atom positions lead to a decrease in the model's energy. After about 50 steps, the energy finally converges. We assume that our model now matches the natural folding of the protein to a sufficient degree.

### 3.2.2 Fragmentation

We run our range of window lengths over the chain's residues in three nested loops.

```
INPUT
- 1 protein with 2 chains: 1st chain with 5 residues, 2nd chain with 7 residues
- Window lengths: 3 - 4
OUTPUT
- 1st chain's fragments:   +++--, -+++-, --+++, ++++-, -++++
- 2nd chain's fragments:   +++----, -+++---, --+++--, ---+++-, ----+++, ++++---, -++++--,
                           --++++-, ---++++
```

*Example 1: The command line parameters* ./programname ./X.pdb x 3 5 *result in the output directory* x *with two subdirectories* x_1, x_2, *each including 3 files* data3.txt, data4.txt, *and* data5.txt. *They contain the results of the computation for the respective fragment length. The selected residues are marked by a „+", the discarded ones by a „-".*

### 3.2.3 SAS Area Computation

The SAS area of the fragments is important with regard to their potential immunogenicity: if a fragment has a very small SAS area, it is buried and hence cannot serve as an epitope in the full protein. If it is mostly solvent exposed, it may constitute an epitope.

Water molecules are not part of a protein, but they have significant influence on a chain's spatial conformation. They are part of PDB files, however, they are unwanted in our computation because they have no biological relevance in this context. Therefore we delete the water after the optimization of the hydrogen atoms and prior to the energy computation. Before we fragment the chain, we assign the according atom radii to all atoms. Computing and storing the SAS area of each atom is done in just one more step. We add up the SAS areas of all atoms of a fragment in the inner nested loop.

### 3.2.4 Secondary Structure Assignment

The secondary structure is of interest in this context, because we want to evaluate whether there is a relation between particular secondary structures prevailing in a fragment and the fragment's energy. If a secondary or supersecondary structure like e.g. helix/turn/helix is associated with low fragment energy, it may be associated with peptide immunogenicity, too.

Each residue is assigned a secondary structure prior to fragmentation. Afterwards in the nested loop we put together a string holding the fragment's secondary structure. This string equates to the final secondary structure output.

| **Secondary Structure:** | Helix | Coil | Turn | Strand | Unknown |
|---|---|---|---|---|---|
| **Letter:** | H | C | T | S | U |

*Table 2: A fragment's secondary structure is described by a string consisting of the letters H, C, T, S, and U. Usually only the letters H, C, and S occur. A fictional fragment with 6 residues might for example result in the describing string „SCCCHH".*

### 3.2.5 Energy Computation

We use the force field AMBER for our energy computations.

| | |
|---|---|
| $E_{total}$ | $= E_{bonded} + E_{nonbonded}$ |
| $E_{bonded}$ | $= E_{bonds} + E_{angles} + E_{torsions}$ |
| $E_{nonbonded}$ | $= E_{electrostatic} + E_{van\ der\ Waals}$ |

*Table 3: AMBER sums terms representing bonds, angles, torsions, electrostatic interactions, and van der Waals forces.*

It brings together bonded terms relating to covalently bonded atoms with nonbonded terms describing long-range interactions, like electrostatic and van der Waals forces [PC03].

$$\sum_{bonds} \frac{1}{2} k_b (l - l_0)^2 \qquad (1)$$

$$\sum_{angles} k_a (\theta - \theta_0)^2 \qquad (2)$$

$$\sum_{torsions} \frac{1}{2} V_n [1 + \cos(n\omega - \gamma)] \qquad (3)$$

$$\sum_{j=1}^{N-1} \sum_{i=j+1}^{N} \left\{ \epsilon_{i,j} \left[ \left(\frac{\sigma_{ij}}{r_{ij}}\right)^{12} - 2 \left(\frac{\sigma_{ij}}{r_{ij}}\right)^6 \right] + \frac{q_i q_j}{4 \pi \epsilon_0 r_{ij}} \right\} \qquad (4)$$

The bonded AMBER term is the sum of the equations (1), (2), and (3).
Equation (1) descrices the forces between covalently bonded atoms by a harmonic potential.
Equation (2) sums over the energies resulting from the geometry of electron orbitals involved in covalent bonding. Twisting a bond due to bond order and neighbouring bonds or lone electron pairs is represented by equation (3).
Equation (4) shows which computations the nonbonded AMBER term arises from: a double summations over i and j takes all atom pairs into account. The first term of the summation represents the electrostatic interactions, the second one stands for the van der Waals forces.

In our implementation the energy computation for a fragment starts when all of its residues have been selected. We set up a force field on this selection and calculate its total energy. The resulting value is the self-energy, i.e. the sum of all energies resulting from the fragment atoms' interactions with other fragment atoms. Interactions with and of the protein's outlying residues are excluded. Finally we remove the selection.

Additionally, we want to know the energies of the fragment's interactions with the rest of the protein. We need a new force field, however this time we set it up on the whole protein. Afterwards we select the same window as before. Then we calculate the total energy of the window's interactions with itself *and* the other protein residues („full energy").

The difference of the energy we computed in the previous step and the self-energy is a term we are strongly interested in. This energy difference stands for an energy term itself: it includes solely the interactions between the fragment atoms and the atoms outside of the examined window. So it describes how stable the fragment would be, if it was not part of the protein. The more the fragment gets stabilized by interactions with the residual protein, the less stable it would be in isolation, and vice versa. This formula quantifies the interrelations:

$$E_{difference} = E_{full} - E_{self}$$

If the self-energy is rather low and much lower than the full energy, the fragment's corresponding peptide is probably relatively stable. The computed value does not equal the folding free energy $\Delta G$, but it is an approximation, and this is what we aimed at developing.

## 3.3 Webserver

We set up a webserver for the analysis of proteins with regard to their fragments' potential immunogenicity.

The interface first allows the user to retrieve a PDB file directly from the RCSB Protein Data Bank by entering a PDB identifier, or to upload a file from his computer.
A minimum and maximum window length has to be chosen, the preselection being 8 or 22 residues, respectively.

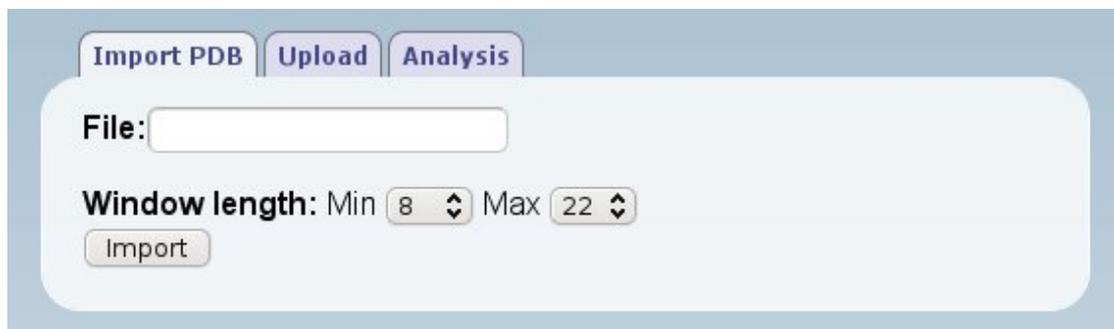

*Figure 7: The tabs allow the user to choose between file import from the PDB and file upload from his computer.*

After importing the file, it is being processed by the C++ program described in chapter *3 Implementation*.

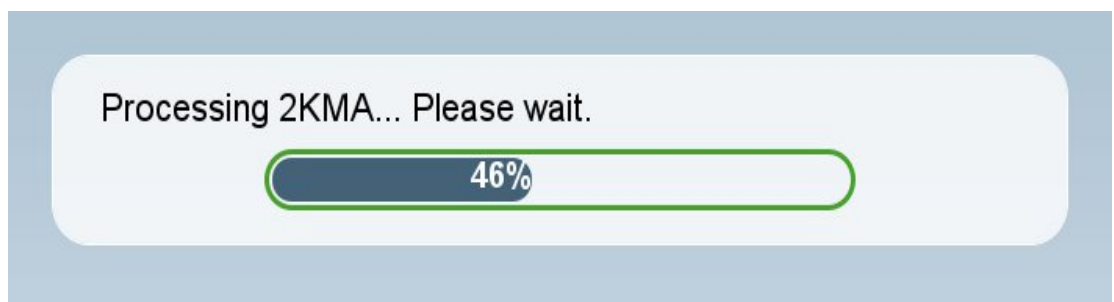

*Figure 8: A progress bar indicates the status of the computation. In this example the C++ program has completed about 18% of the analysis for the PDB file 2KMA.*

After the processing is completed, our server provides the results for each protein chain. The user is being offered a choice of evaluations.

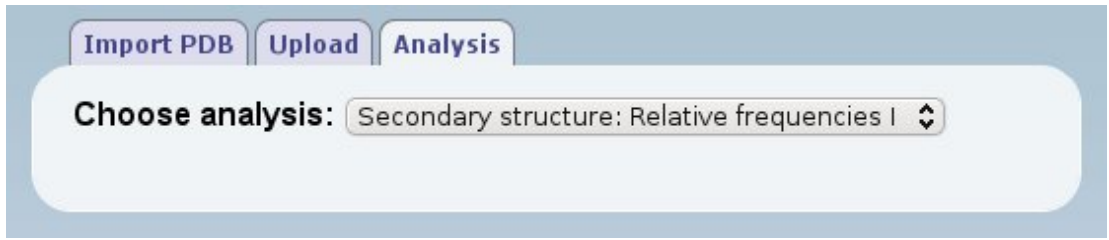

*Figure 9: The drop-down menu makes the implemented analyses available to the user. In this example the user chose an analysis based on the relative frequencies of secondary structure elements compared to the fragments' energy difference.*

The available options are:
1. Secondary structure: Absolute frequencies
2. Secondary structure: Relative frequencies I
3. Secondary structure: Relative frequencies II
4. Secondary structure: High energy difference
5. Solvent accessible surface area

Examples explain their meaning („Toggle examples"). Each analysis offers two download options: the results can be exported for printing in Portable Document Format (PDF) or for direct data access in Extensible Markup Language (XML). Exporting to XML makes the re-use of analysis results in other software applications especially easy.

Additionally the user interface features a visualization of the complete protein structure imported from the PDB.
Interactive charts for each window length visualize the fragments' self-energy and energy difference function, SAS area, and secondary structure distribution. If the user wants to save the chart to his computer, he can do so by clicking the button *„Convert to image"*. Thereupon the interactive chart is being replaced by a graphic image, which can be saved for later usage.
The service will be made available to the public.

# 4 Results and Discussion

## 4.1 General Properties of our Model

First we wanted to know whether our program's results were comparable to Camacho *et al.*'s results, so we let it compute the energy values for HRS. To obtain useful results, we based our analysis on the model also used by Camacho.

Not only did we examine HRS, but the self-energy of many other proteins, too. In order to achieve relevant results, we evaluated the non-homologous protein domains from the Nh3D 3.0 reference dataset. We deleted particular proteins by hand, so that the remaining dataset guarantees that no pairwise sequence identities after global alignment are greater than 30%. All validation results show an elementary amount of similarity in certain aspects.

Also there are some potential error sources in our energy model. We will discuss them in this chapter.

### 4.1.1 Energy and RMSD

First of all we compared our results with the findings of Camacho *et al.* in a line chart. Unfortunately, Camacho examined only 10 HRS fragments. This is a very small set for a meaningful comparison.

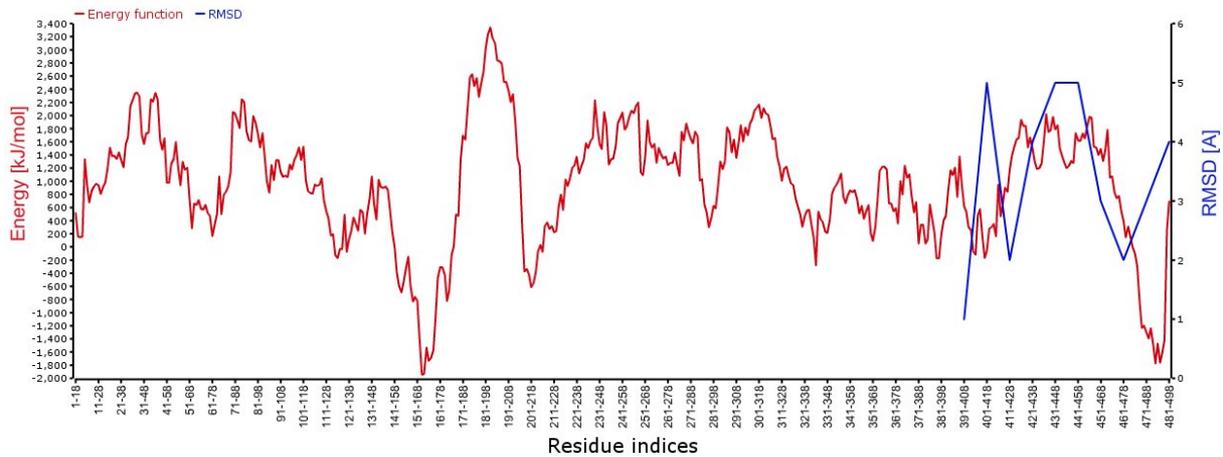

*Figure 10: The chart makes comparable the self-energy (red) to the RMSD values (blue).*

Upon visual inspection, the curves' progressions between the 6 fragments 411-428 to 461-478 show a rather strong correlation, while for the 3 remaining fragments there is not such an obvious relationship between self-energy and RMSD.

Our energy computation is only a model – a fact that applies to the RMSD calculation as well. Therefore neither our results really describe the proteins in their natural way, nor do those of Camacho *et al.*

We analyzed proteins with our model, later on we will let them undergo laboratory tests uncovering their real immunogenicity. Unlike this chapter's comparison, this is a reliable way to check the biological significance of our model's results.

## 4.1.2 Energy and Secondary Structure

We are interested in the interrelation between the energy functions and a fragment's secondary structure. The results show a correlation between self-energy and helix. Since the self-energy and the energy difference are correlated negatively, there is a negative correlation between energy difference and helix, too. For strand and coil it is just the other way around.

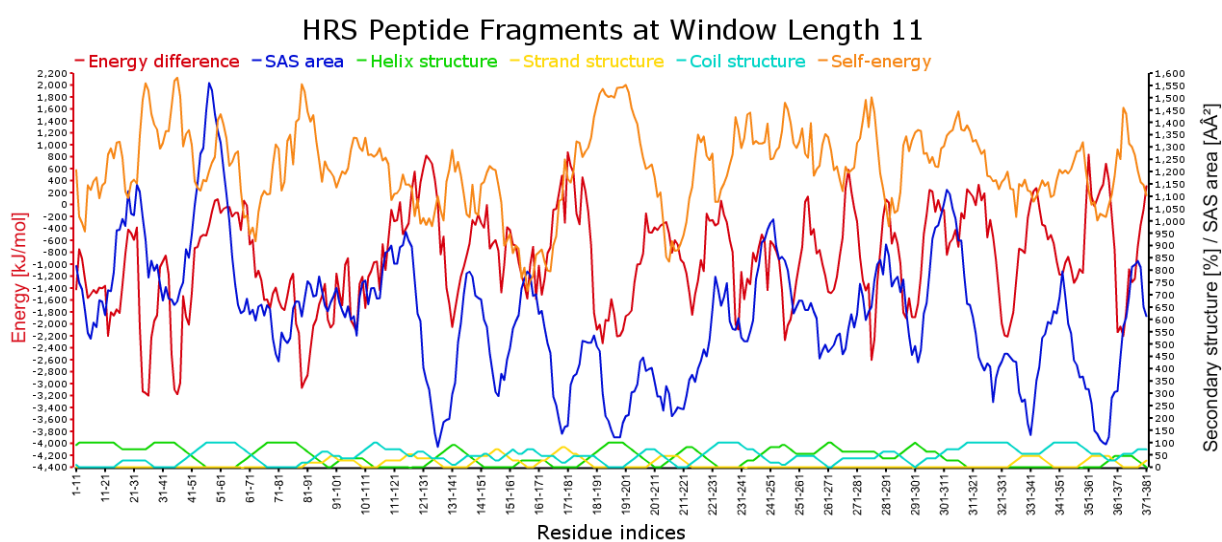

*Figure 11: Self-energy (orange) correlates with the curve indicating helices (green), while it correlates negatively with the energy difference (red) in this example of HRS fragments with 11 residues.*

However, we do not conclude that helices are energetically unfavorable, since a helix is known to be a common secondary structure with low energy resulting from its hydrogen bonds. Helix is expected to be the structure with the highst stability, because our fragments include its stabilizing hydrogen bonds, while the bonds stabilizing a beta strand often come from a second parallel or antiparallel strand.

We assume that the surprising results may be caused by a weakness of our model. We discuss this possibility in chapter *4.1.5 Shortcomings of our Model*.

Beyond that we studied the occurence of the supersecondary structures *hairpin* (strand/coil/strand) and *helix/coil/helix*. Our definition requires that a hairpin consist of at least two residues in strand structure followed by at least one residue in coil structure followed again by at least two residues in strand structure. The same scheme applies to helix/coil/helix.

| Window length | Coil | Helix | Strand | Coil & Helix | Coil & Strand | Helix & Strand | Coil, Helix & Strand | Hairpin | Helix/Coil /Helix |
|---|---|---|---|---|---|---|---|---|---|
| 8 | 40.58% | 1.32% | -1% | 24.16% | 59.09% | 22.22% | 48.98% | 17.65% | 28.57% |
| 9 | 41.67% | 0% | -1% | 21.39% | 60.38% | 25% | 50% | 27.27% | 32.14% |
| 10 | 47.17% | 0% | -1% | 18.97% | 60% | 28.57% | 48.75% | 25.93% | 31.43% |
| 11 | 47.83% | 0% | -1% | 17.82% | 63.83% | 20% | 44.33% | 34.38% | 30.95% |
| 12 | 55% | 0% | -1% | 19.02% | 61.8% | 0% | 40.71% | 35.14% | 30% |
| 13 | 51.43% | 0% | -1% | 19.9% | 61.45% | 0% | 40.31% | 30.95% | 31.03% |
| 14 | 54.84% | 0% | -1% | 19.02% | 61.04% | 0% | 40% | 38.3% | 27.27% |
| 15 | 55.56% | 0% | -1% | 19.61% | 59.72% | -1% | 39.62% | 44.23% | 25.68% |
| 16 | 52.17% | 5.56% | -1% | 20.69% | 59.7% | -1% | 38.37% | 47.37% | 21.95% |
| 17 | 47.37% | 14.29% | -1% | 20.79% | 53.23% | -1% | 40% | 50% | 17.78% |
| 18 | 53.33% | 27.27% | -1% | 20.79% | 50.88% | -1% | 39.8% | 50.75% | 13.27% |
| 19 | 54.55% | 25% | -1% | 18.81% | 53.85% | -1% | 41.55% | 54.17% | 9.62% |
| 20 | 50% | 16.67% | -1% | 21% | 46.81% | -1% | 41.28% | 52.63% | 11.82% |
| 21 | 33.33% | 0% | -1% | 20.6% | 44.19% | -1% | 42.92% | 50.62% | 12.07% |
| 22 | 0% | 50% | -1% | 22.22% | 33.33% | -1% | 43.16% | 47.67% | 13.93% |

*Table 12: The expected value for the occurence of any structure in fragments with hihg energy difference is 33.33%. While helix/coil/helix tends to occur in fragments with low energy difference, hairpin often is found in those with high energy difference.*

This tabular analysis provides insight into the distribution of secondary structures over fragments with high energy difference. The consequence of helices being associated with energy difference levels is that helix/coil/helix hardly occurs in those fragments. Hairpin seems to be a more common structure in this kind of fragments.

However, the significance of these numbers depends on the quality of our model and are influenced by the afore mentioned shortcomings.

### 4.1.3 Energy and SAS Area

An epitope must lie on the surface of its molecule in order to be able to interact with antibodies. Thus, a particularly small SAS area is a knock-out criterion for fragment immunogenicitiy in this context.

Aside from this consideration, the SAS area of a fragment also reveals a minor interrelation with fragment energy.
Folded proteins usually have a hydrophobic core in which side chain packing stabilizes the folded state, and charged or polar side chains occupy the surface where they interact with surrounding water. Thus, the self-energy terms should correlate at least weakly negatively with the SAS area, which is what they actually do – with the exception of small proteins where nearly no fragments are buried.

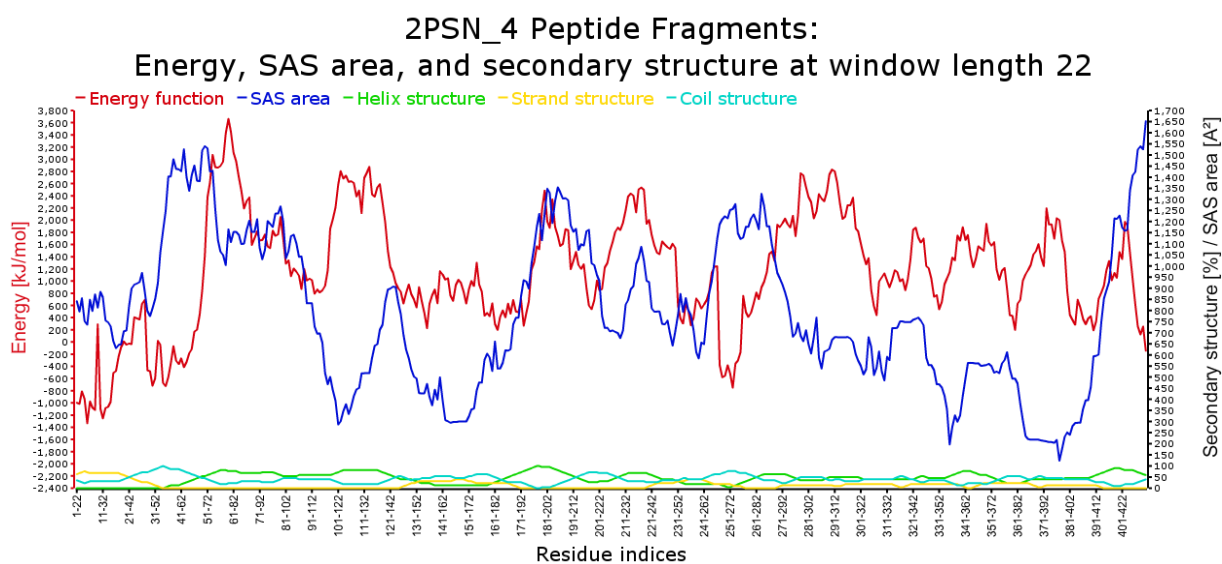

*Figure 13: The self-energy (red) and SAS area curve (blue) show a weak negative correlation.*

### 4.1.4 Smoothness

When comparing the charts for two protein fragments with different window lengths, it can be seen that longer fragments yield more smooth curves. When computing longer fragments, the overlap of two consecutive fragments increases, too. Thus, the influence of the variable residue's self-energy, SAS area, and secondary structure terms on the fragment's properties is smaller the more identical residues two fragments have.

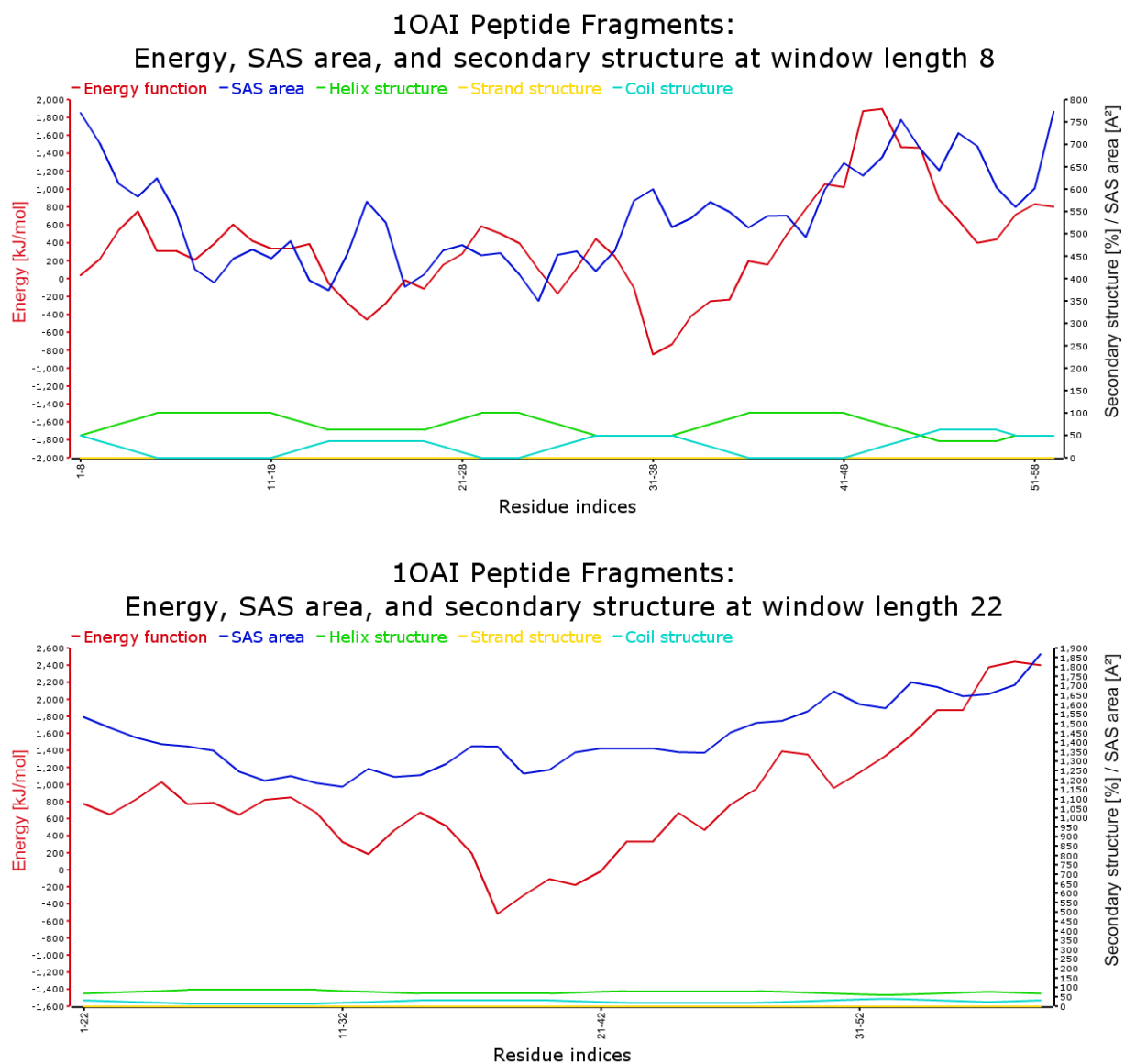

*Figure 14: 1OAI, randomly picked from the reference dataset, shows the curves' smoothing.*

### 4.1.5 Shortcomings of our Model

Some aspects of protein structure are not fully taken into account by our model.

The amino acid cystein has a thiol side chain, which often is involved in disulfide bonds. These covalent bonds stabilize proteins' tertiary and quarternary structures, but our model does not provide a special handling for them.

The same applies for ligands. We delete them prior to the energy computation, but they also contribute to the overall protein energy and stability.

Also, we do not pay full attention to the fact that the two amino acids at the ends of the fragment in most cases are covalently bonded to outlying residues. Maybe it is necessary to break these bonds prior to the energy computation. Breaking the peptide bonds would require an additional treatment of the N-terminal and C-terminal ends of the fragment. Adding caps (-H, respectively -OH) could solve this problem.

Finally, the high energy values for helices could be a result of unsufficient hydrogen atom position optimization. Splitting up the self-energy term revealed that the high energy level is a result of extraordinarily strong electrostatic interactions in fragments with helices. This is probably a weakness of our model.

## 4.2 Protein Analyses

Though we examined many proteins in order to obtain a broad dataset for evaluation, we were particularly interested in the findings about some of them. Besides HRS we took a closer look at Enolase I (ENO1). It will later be subject to laboratory examinations since matching monoclonal antibodies are available.

### 4.2.1 HRS

HRS is a medium sized protein consisting of one chain only. Since it is known to have implications as an autoantigen, it is an interesting subject of study.
Our evaluations show the curves' increasing smoothness, the longer the fragments are. There is a slight negative correlation between self-energy and SAS area.

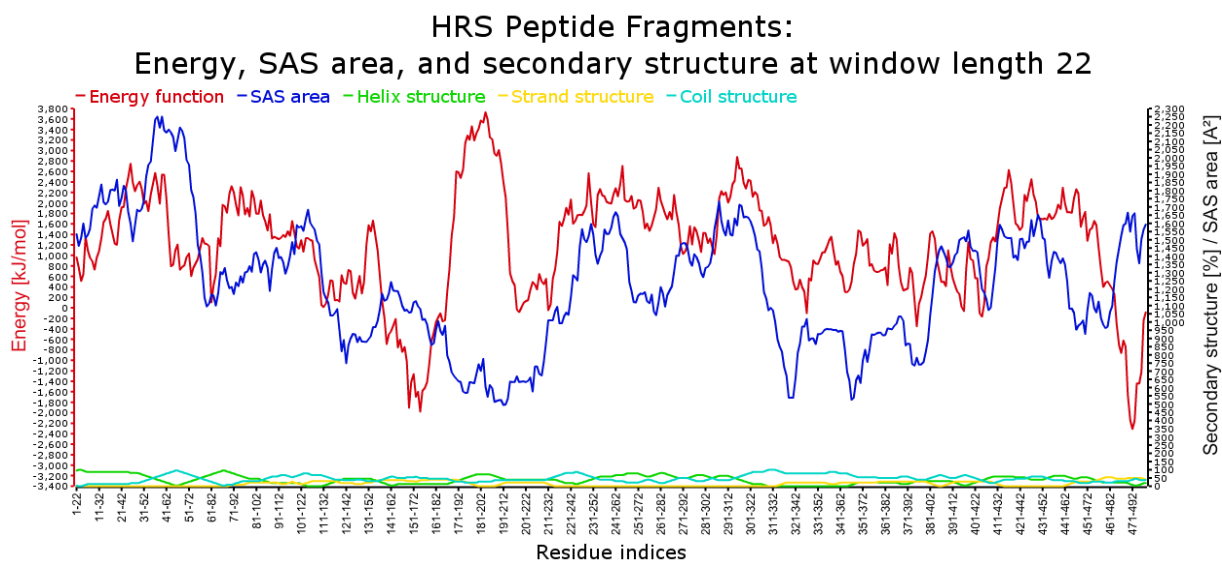

*Figure 15: The chart for HRS shows all typical characteristics of our model.*

With regard to biological sampling we recommend fragments with certain properties: not too small SAS area, low self-energy and higher energy difference. In the case of HRS we picked out the fragments between residue 138 and 147, as well as those between 463 and 498. Window length does not apply here, since fragments with varying residue count should be subject to examination.

These recommendations should be taken with reserve since they are based on results possibly biased by wrong helix energies.

### 4.2.2 ENO1

The human lyase enolase 1 alpha consists of 4 chains, out of which only one is sequence-unique: chain A. The PDB file comes with two ligands included. One of them is not being deleted successfully by our C++ program, since in this case the PDB file was not standards-compliant. Therefore we deleted it by hand.

Our charts reveal that there is little difference between chain A and the other chains.

Because of the very long runtime required for the energy difference computation of this relatively large protein, we only evaluated its self-energy.

The curves of helix and self-energy correlate negatively, as they did in the case of HRS, and so do self-energy and SAS area curve, but in a less obvious manner. The smoothness increases when chosing larger fragment visualizations.

The fragments recommended for further immunological investigations can be found between residues 33 and 70, as well as between 248 and 286.

These recommendations should be taken with reserve, too.

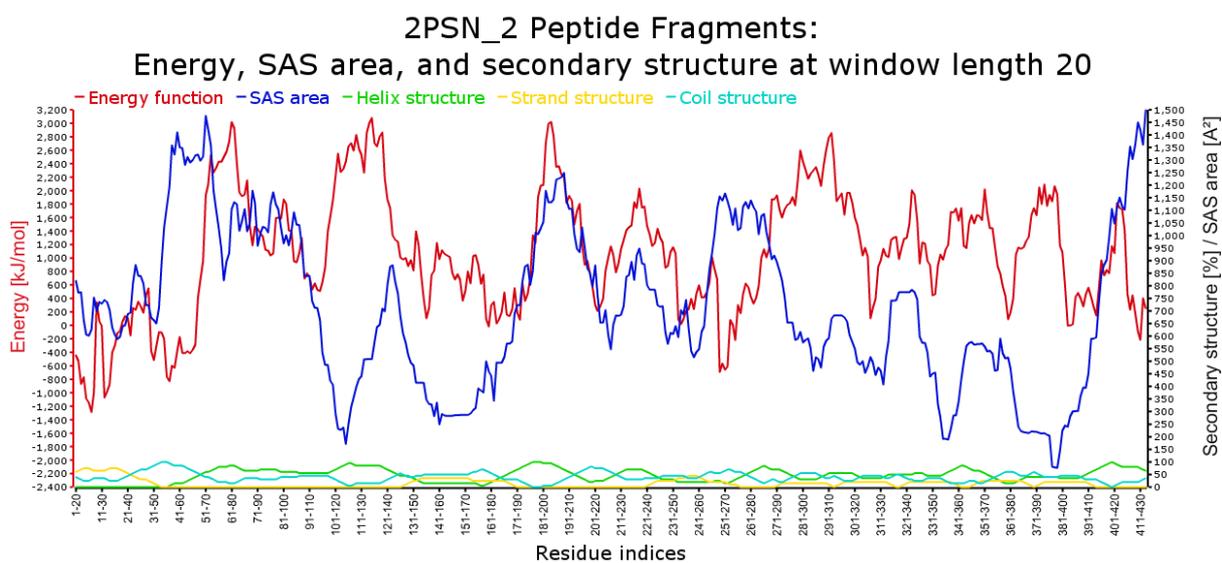

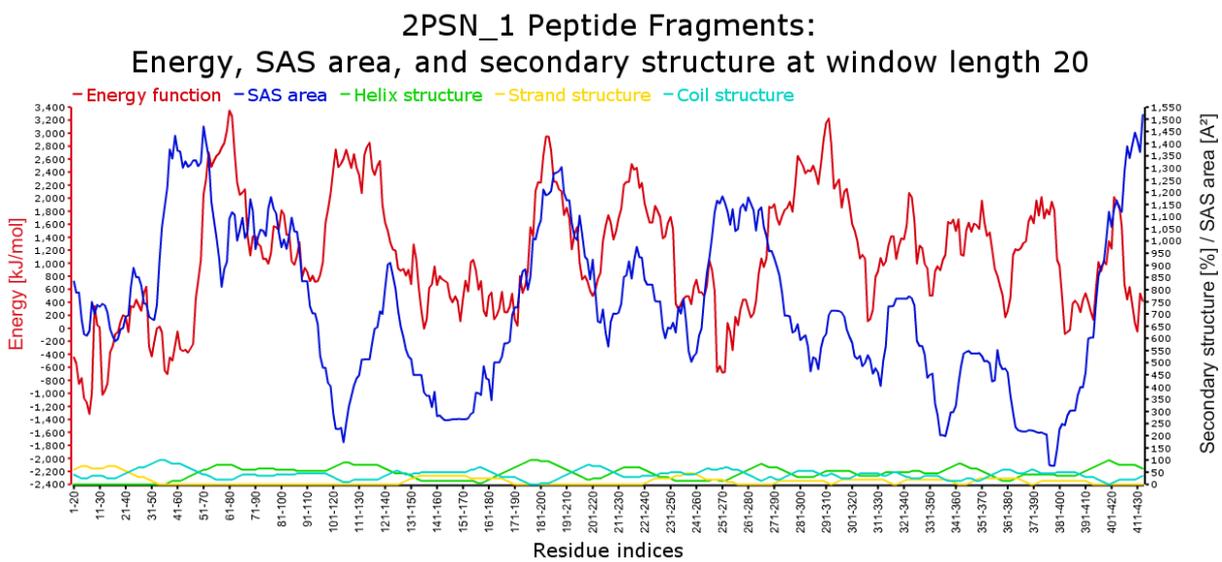

*Figure 16: Although ENO1's chain A (2PSN_1) and chain B (2PSN_2, see above) have slightly different sequences, their overall appearance is rather similar.*

# 5  Conclusion and Outlook

This thesis focuses on the development of an energy model linked to peptide stability.

In the previous chapters we introduced the fundamental concepts of folding stability, immunogenicity, and the interrelation of both. We presented the implementation and validation of our suggested model, which is based on the stability of corresponding protein fragments. The explanation of the validation's results exposed the potential, as well as the inherent difficulties of our approach.

There are many ways how the output of this thesis can still be improved, extended, and used further in the future.

The C++ program itself can perhaps be sped up by optimizing the code.

The implementation of the energy model could by modified in order to yield more realistic results, especially for helices.

Furthermore it is possible to take implicit solvation by the surrounding medium into account in our model, and maybe to even carry out free energy perturbation (FEP) calculations.

In order to find a better start conformation for the fragments one could also sample the energy space, e.g. with a simulated annealing algorithm.

The webserver could be improved by letting a database do the data and user management. The currently implemented method, which is based on text files, is rather simplistic. A molecular visualization with a colouring according to fragment energy could be integrated using BALLView, a molecular modeling and visualization application [MHL+05].

Previous to carrying out laboratory tests with the afore mentioned fragments, they could be synthesized and analyzed via nuclear magnetic resonance (NMR) spectroscopy in order to gain physical data about the stability of the molecules. The NMR spectroscopy technique offers the possibility to reveal the structure of solute molecules. Depending on the results' degree of resolution one can make conclusions about the stability of the peptide. A high resolution is an indicator for high stability. Peptides leading to a low resolution need not be subject to further tests.

The actual immunogenicity of eligible candidates could be tested with tissues and/or model organisms, e.g. mice (Mus musculus) or rats (Rattus norvegicus).

In this context it would be interesting to take a closer look at peptides with mutations. Changes in the peptide sequence can change the molecule's stability. One could investigate which mutations increase stability and are known autoimmunogens, too. A starting point for searching mutations involved in a disease is the Roche Cancer Genome Database (RCGDB) [KEL+10]. This data pool has been released recently and is freely available online.

# Table of Abbreviations

| | |
|---|---|
| AMBER | Assisted Model Building with Energy Refinement |
| APC | Antigen-presenting cell |
| BALL | Biochemical Algorithms Library |
| BCR | B cell receptor |
| ENO1 | Enolase 1 alpha |
| FEP | Free energy perturbation |
| HRS | Histidyl-tRNA synthetase |
| MD | Molecular dynamics |
| MHC | Major histocompatibility complex |
| NMR | Nuclear magnetic resonance |
| PDB | Protein Data Bank |
| PDF | Portable Document Format |
| RCGDB | Roche Cancer Genome Database |
| RMSD | Root mean square deviation |
| SAS | Solvent accessible surface |
| XML | Extensible Markup Language |